\documentclass{aastex61}

\date{\today}

\begin{document}

\title{Planck 2015 constraints on the non-flat $\Lambda$CDM inflation model}

\author{Junpei Ooba}
\altaffiliation{ooba.jiyunpei@f.mbox.nagoya-u.ac.jp}
\affiliation{Department of Physics and Astrophysics, Nagoya University, Nagoya 464-8602, Japan}

\author{Bharat Ratra}
\affiliation{Department of Physics, Kansas State University, 116 Cardwell Hall, Manhattan, KS 66506, USA}

\author{Naoshi Sugiyama}
\affiliation{Department of Physics and Astrophysics, Nagoya University, Nagoya 464-8602, Japan}
\affiliation{Kobayashi-Maskawa Institute for the Origin of Particles and the Universe, Nagoya University, Nagoya, 464-8602, Japan}
\affiliation{Kavli Institute for the Physics and Mathematics of the Universe (Kavli IPMU), The University of Tokyo, Chiba 277-8582, Japan}



\begin{abstract}

We study Planck 2015 cosmic microwave background (CMB) anisotropy data using 
the energy density inhomogeneity power spectrum generated by quantum 
fluctuations during an early epoch of inflation in the non-flat $\Lambda$CDM model.
Unlike earlier analyses of non-flat models, which assumed an 
inconsistent power-law power spectrum of energy density inhomogeneities,
we find that the Planck 2015 data alone, and also in conjunction with baryon acoustic oscillation measurements,
are reasonably well fit by a closed $\Lambda$CDM model in which spatial curvature contributes
a few percent of the current cosmological energy density budget. In this model, the measured Hubble 
constant and non-relativistic matter density parameter are in good agreement 
with values determined using most other data. Depending on parameter values, 
the closed $\Lambda$CDM model has reduced power, relative to the tilted, 
spatially-flat $\Lambda$CDM case, and can partially alleviate the low 
multipole CMB temperature anisotropy deficit and can help partially reconcile 
the CMB anisotropy and weak lensing $\sigma_8$ constraints, at the expense of 
somewhat worsening the fit to higher multipole CMB temperature anisotropy data.
Our results are interesting but tentative; a more thorough analysis is needed to properly gauge their significance. 

\end{abstract}

\keywords{cosmic background radiation --- cosmological parameters --- large-scale structure of universe  --- inflation --- observations}



\section{Introduction} \label{sec:intro}

In the standard spatially-flat $\Lambda$CDM cosmological scenario 
\citep{Peebles1984}, the currently accelerating cosmological expansion is 
powered by a cosmological constant $\Lambda$ that dominates the current 
cosmological energy budget. Cold dark matter (CDM) and baryonic matter are the 
second and third biggest terms in the energy budget, followed by small 
contributions from neutrinos and photons. In this model spatial hypersurfaces 
are taken to be flat. For reviews of this scenario, as well as of the dynamical 
dark energy and modified gravity pictures, see  \citet{RatraVogeley2008}, 
\citet{Martin2012}, \citet{Joyceetal2016}, and references therein.

The presently popular version of the standard flat-$\Lambda$CDM model is 
conveniently parameterized in terms of six variables: the current values of 
the baryonic and cold dark matter density parameters
multiplied by the square of the Hubble constant (in units of 100 km s$^{-1}$ 
Mpc$^{-1}$), $\Omega_{\rm b} h^2$ and $\Omega_{\rm c} h^2$; the angular diameter 
distance as a multiple of the sound horizon at recombination, $\theta$; the 
reionization optical depth, $\tau$; and the amplitude and spectral index of the
(assumed) power-law primordial scalar energy density inhomogeneity power 
spectrum, $A_{\rm s}$ and $n_{\rm s}$, \citep{Adeetal2016a}. The predictions of 
the flat-$\Lambda$CDM model appear to be in reasonable accord with most 
available observational constraints 
\citep[][and references therein]{Adeetal2016a}.    

However, there are some suggestions that flat-$\Lambda$CDM might not be as compatible with measurements of the Hubble parameter
\citep{Sahnietal2014, Dingetal2015, Zhengetal2016}. Also, flat-$\Lambda$CDM might be less favored by 
a combination of measurements \citep{Solaetal2017a, Solaetal2017b, Solaetal2018, Solaetal2017c, Zhangetal2017} that might be better fit by dynamical dark 
energy models, including the simplest, physically consistent, seven parameter 
flat-$\phi$CDM model in which a scalar field 
$\phi$ with potential energy density $V(\phi) \propto \phi^{-\alpha}$, $\alpha
> 0$, is the dynamical dark energy \citep{PeeblesRatra1988, RatraPeebles1988} 
and $\alpha$ is the seventh parameter and governs dark energy evolution. If 
future data strengthens these results, this will be an important pointer 
towards an improved cosmological model.  

On the other hand, the spatial curvature energy density term in the Friedmann
equation is also dynamical, albeit one that evolves faster than an acceptable
dynamical dark energy contribution. It is therefore possible that a non-flat 
$\Lambda$CDM model might also provide a reasonable fit to the data of the 
previous paragraph that indicate evidence for dynamical dark energy. It is 
often suggested that cosmic microwave background (CMB) anisotropy data 
\citep[][and references therein]{Adeetal2016a} demand spatially-flat hypersurfaces.
However the latest Planck 2015 examination of this issue was 
based on an energy density perturbation power spectrum that is 
only appropriate for flat spatial hypersurfaces \citep{Adeetal2016a};
spatial curvature sets an additional length scale so the power-law power spectrum 
assumed in that analysis is not physically consistent.

Inflation provides a way of computing the power spectrum in the non-flat model.
For open spatial hypersurfaces it is usual to assume a \citet{Gott1982} 
open-bubble inflation model as the initial epoch of the cosmological model. 
One then computes zero-point quantum fluctuations during open inflation and 
propagates these to the current open accelerating universe where they are 
energy density inhomogeneities \citep{RatraPeebles1994, RatraPeebles1995, Bucheretal1995, LythWoszczyna1995, Yamamotoetal1995}.\footnote{For some observational consequences 
of the open inflation power spectrum see \citet{Kamionkowskietal1994}, 
\citet{Gorskietal1995}, \citet{Gorskietal1998}, \citet{Ratraetal1999}, and 
references therein.}
Unlike in the flat case, this power spectrum 
is not a power law in wavenumber, but it is the generalization to 
the open model \citep{LythStewart1990} of the flat case scale-invariant 
spectrum \citep{Harrison1970, PeeblesYu1970, Zeldovich1972}.

One can use Hawking's prescription for the quantum state of the universe \citep{Hawking1984}
to construct a closed inflation model 
\citep{Ratra1985, Linde1992, LindeMezhlumian1995, Grattonetal2002, Linde2003, LasenbyDoran2005, Ratra2017}.\footnote{For discussions of a different closed 
inflation model see \citet{Ellisetal2002a}, \citet{Ellisetal2002b}, and 
\citet{Uzanetal2003}.}
Here the constants of integration in the closed inflation epoch 
linear perturbation solutions are determined from closed de Sitter 
invariant quantum mechanical initial conditions in the Lorentzian section 
of the closed de Sitter space that are derived from Hawking's prescription 
that the quantum state of the universe only include field configurations 
that are regular on the Euclidean (de Sitter) sphere sections \citep{Ratra1985, Ratra2017}.   
Zero-point quantum-mechanical fluctuations during closed inflation 
provide a late-time energy density inhomogeneity power spectrum that is not 
a power law in wavenumber \citep{Ratra2017} but is a generalization to the 
closed case \citep{WhiteScott1996, Starobinsky1996, Zaldarriagaetal1998, Lewisetal2000, LesgourguesTram2014} of the flat-space scale-invariant 
spectrum.\footnote{Other power spectra, that apparently differ from what we 
use here, have been considered in the closed model \citep{Efstathiou2003, LasenbyDoran2005, Massoetal2008, Bongaetal2016}. As discussed in \citet{Ratra2017}, 
the power spectrum we use here is based on closed de Sitter invariance during 
inflation and it is unclear how to interpret any other power spectrum.} 

In both the open and closed cases, there is no simple tilt option,
so $n_{\rm s}$ is no longer a free parameter and is replaced by the current value 
of the curvature density parameter $\Omega_{\rm k}$ which results in a six 
parameter non-flat $\Lambda$CDM inflation model. This is a physically 
consistent non-flat model that can be used for analyses of CMB
anisotropy and other measurements. If needed, one may generalize this model to 
a seven or more parameter model that allows for dynamical scalar field 
(or other) dark energy \citep{Pavlovetal2013}.   

In this paper we utilize the Planck 2015 CMB anisotropy data to constrain this 
six parameter non-flat $\Lambda$CDM model. This is the first time the Planck 2015 CMB anisotropy data have
been studied in the context of a physically 
consistent non-flat model. Unlike \cite{Adeetal2016a} who used a seven parameter
non-flat model in their derivation of 
limits on spatial curvature, here we find in our simpler six parameter non-flat 
model that neither do the CMB anisotropy data alone, nor do the CMB 
anisotropy data in conjunction with baryon acoustic oscillation (BAO) 
measurements demand that spatial hypersurfaces be flat.
In fact, the data favor a mildly closed model.

In our analyses here we use a number of CMB anisotropy data combinations
\citep{Adeetal2016a}. For CMB data alone, we find that the best-fit non-flat
$\Lambda$CDM model, for the TT + lowP + lensing Planck 2015 data, has spatial 
curvature density parameter $\Omega_{\rm k} = - 0.018 {+ 0.010 \atop - 0.007} {+0.018 \atop - 0.020}$ (1 and 2$\sigma$ error bars) and is slightly closed. 
For comparison, for the power-law power spectrum \citet{Adeetal2016a}, their 
eqn.\ (49), find their best-fitting seven parameter non-flat model has   
$\Omega_{\rm k} = - 0.005 {+0.016 \atop - 0.017}$ (2$\sigma$ error bars) and 
is consistent with having flat spatial hypersurfaces. When we include the 
same BAO data that Planck 2015 used, we find for the TT + lowP + lensing CMB 
anisotropy case that $\Omega_{\rm k} = - 0.008 \pm 0.002 \pm 0.004$, while 
\citet{Adeetal2016a}, their eqn.\ (50), found   
$\Omega_{\rm k} =  0.000 \pm 0.005$ (2$\sigma$ error bars).

It might be significant
that the best-fit six parameter closed $\Lambda$CDM 
models have less CMB temperature anisotropy $C_\ell$ power at low $\ell$ than
does the best-fit six parameter tilted, spatially-flat $\Lambda$CDM model, 
and so 
are in slightly better agreement with the low-$\ell$ temperature $C_\ell$ 
measurements (less so when the BAO data are included in the mix).\footnote{
There has been much discussion of the lower-$\ell$ temperature anisotropy 
$C_\ell$ power excess predicted in the best-fit spatially-flat tilted 
$\Lambda$CDM model compared to what is observed in the data \citep[][and references therein]{Adeetal2016b, HuntSarkar2015, Schwarzetal2016}, including 
earlier WMAP and COBE data \citep[][and references therein]{Hinshawetal2013}.} 
Because the spatial curvature scale is very large one expects mild 
non-flatness to more significantly affect the lower-$\ell$ $C_\ell$.
The low-$\ell$ improvement comes with
a modest worsening of the fit
to the higher-$\ell$  temperature $C_\ell$ observations. 

The closed $\Lambda$CDM model partial alleviation of the low-$\ell$ $C_\ell$ 
power deficit also results in a slight reduction of $\sigma_8$ in this model 
(for the CMB data alone case). This might be helpful in reducing the 
disagreement between power estimated on this scale from Planck 2015 data and 
(lower) estimates from weak lensing and galaxy cluster observations.
It is also interesting
--- and it might prove significant ---
that the Hubble constant $H_0$
and nonrelativistic matter density parameter $\Omega_{\rm m}$ 
values for the best-fit closed $\Lambda$CDM models are quite reasonable (for 
the CMB alone data this is for the case when CMB lensing is included) and 
very consistent with estimates of these parameters from most other data.      

The structure of our paper is as follows. In Sec.\ II we summarize the methods
we use in our analyses here. Our parameter constraints are tabulated, plotted, 
and discussed in Sec.\ III, where we also attempt to judge how well the 
best-fit closed-$\Lambda$CDM model fits the data. We conclude in Sec.\ IV.

\section{Methods}

\begin{figure}[ht]
\plotone{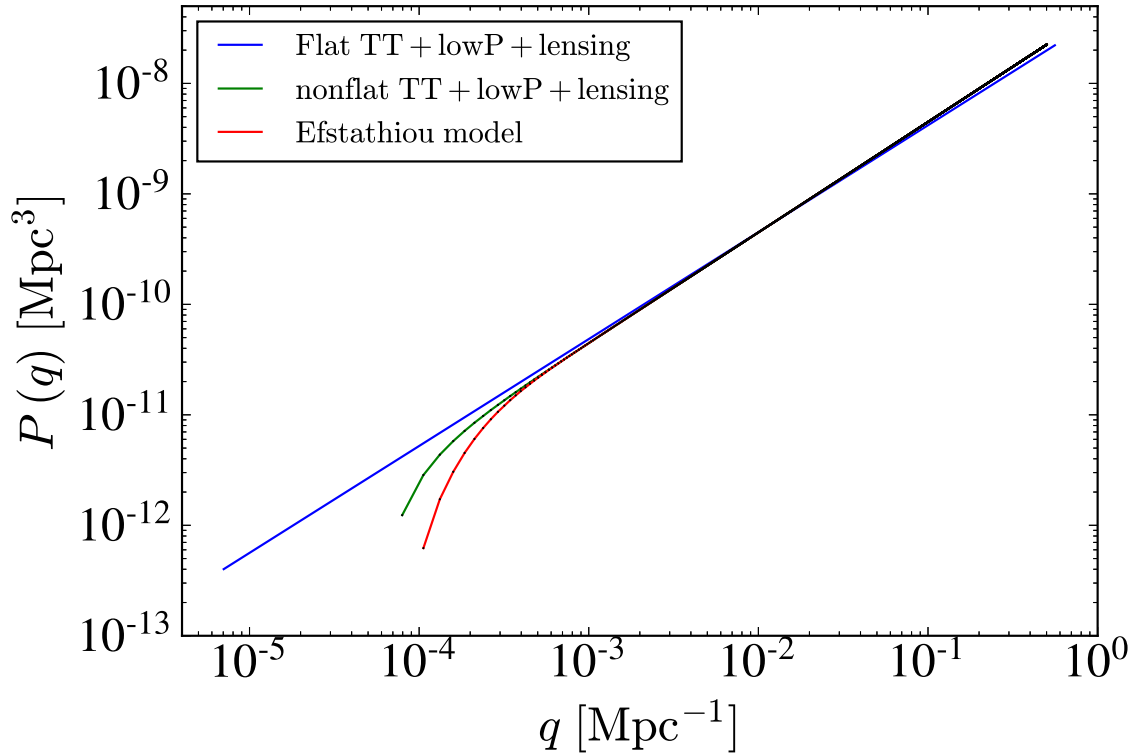}
\caption{Best-fit (TT + lowP + lensing) gauge-invariant fractional energy density inhomogeneity power spectra.
The blue line corresponds to the tilted flat-$\Lambda$CDM model of \citet{Adeetal2016a}.
In the closed case, wavenumber $q \propto A + 1$
where the eigenvalue of the spatial Laplacian is $-A(A+2)$, $A$ is a non-negative integer,
with $A = 0$ corresponding to the constant zero-mode on the three sphere, the power 
spectrum vanishes at $A = 1$, and the points on the green curve correspond 
to $A = 2, 3, 4, ...$, see eqns.\ (8) and (203) of \citet{Ratra2017}.
The red curve shows another closed, but non-inflationary, model power 
spectrum \citep{Efstathiou2003}.
On large scales the power spectrum for the best-fit closed-$\Lambda$CDM 
inflation model 
is suppressed relative to that of the best-fit tilted flat-$\Lambda$CDM 
inflation model.
$P(q)$ is normalized by the best-fit value of $A_{\rm s}$ at the pivot scale $k_0 = 0.05$ [Mpc${}^{-1}$]. We note that the power spectrum of the seven 
parameter best-fit tilted closed-$\Lambda$CDM non-inflationary model of 
\citet{Adeetal2016a} is very close to that shown by the blue line in this plot.
\label{fig:Pk}}
\end{figure}

For our non-flat $\Lambda$CDM model analyses here we use the open and closed 
inflation model quantum energy density inhomogeneity power spectrum
\citep{RatraPeebles1995, Ratra2017}. Figure \ref{fig:Pk} compares a 
closed-$\Lambda$CDM inflation model power spectrum and a tilted 
flat-$\Lambda$CDM inflation model power spectrum.
We use the public numerical code CLASS \citep{Blasetal2011}
to compute the angular power spectra of the CMB 
temperature, polarization, and lensing potential anisotropies. Our parameter 
estimations are carried out using Monte Python \citep{Audrenetal2013} that is
based on the Markov chain Monte Carlo (MCMC) method.

We set flat priors for the cosmological parameters over the ranges
\begin{eqnarray}
\label{eq:prior}
&100\theta \in (0.5,10),\ \ \Omega_{\rm b}h^2 \in (0.005,0.04),\ \ \Omega_{\rm c}h^2 \in (0.01,0.5), \nonumber\\
&\tau \in (0.005,0.5),\ \ {\rm ln}(10^{10}A_{\rm s}) \in (0.5,10),\ \ \Omega_{\rm k} \in (-0.5, 0.5).
\end{eqnarray}
The CMB temperature and the effective number of neutrinos were set
to $T_{\rm CMB}= 2.7255\ \rm K$ from COBE \citep{Fixsen2009} and $N_{\rm
eff}=3.046$ with one massive (0.06 eV) and two massless neutrino species 
in a normal hierarchy. The primordial helium fraction $Y_{\rm He}$ is 
inferred from standard Big Bang nucleosynthesis, as a function of the baryon 
density.

We compare our results with the CMB angular power spectrum data from the
Planck 2015 mission \citep{Adeetal2016a} and the BAO measurements from the 
matter power spectra obtained by the 6dF Galaxy Survey (6dFGS) 
\citep{Beutleretal2011}, the Baryon Oscillation Spectroscopic Survey 
(BOSS; LOWZ and CMASS) \citep{Andersonetal2014}, and the Sloan Digital Sky 
Survey (SDSS) main galaxy sample (MGS) \citep{Rossetal2015}.

\section{Results}

\begin{figure}[ht]
\plottwo{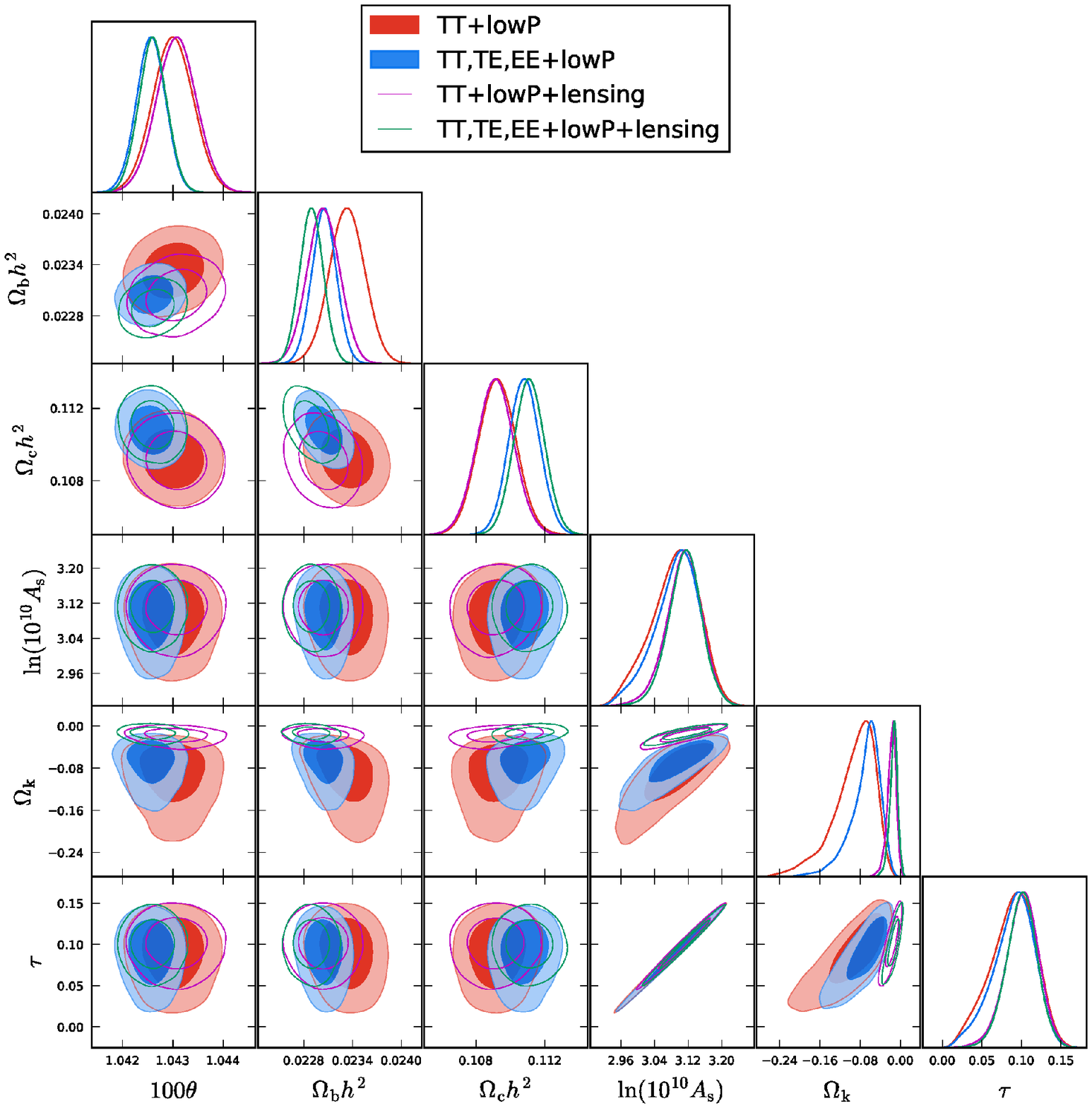}{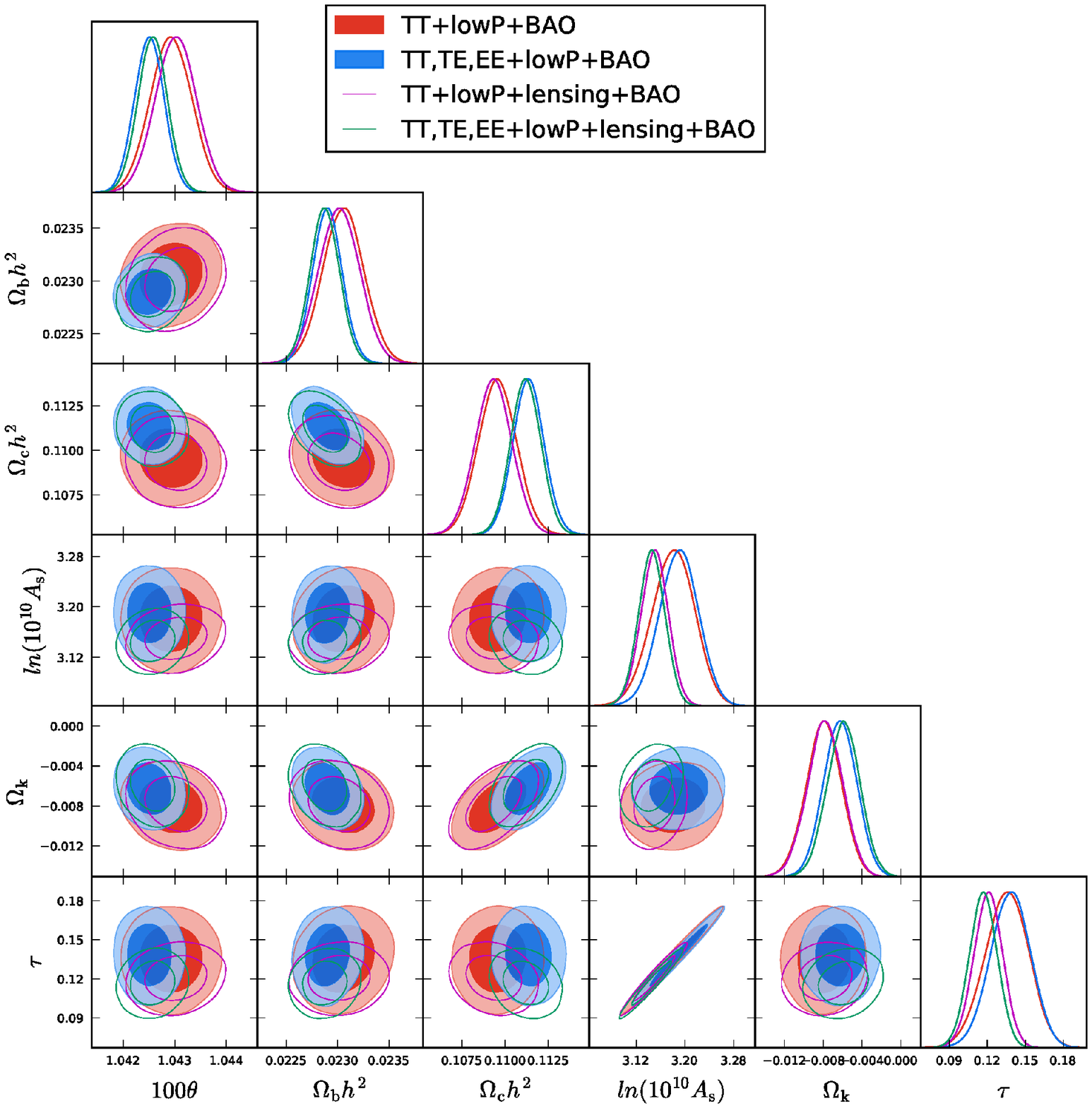}
\caption{$68.27\%$ and $95.45\%$ confidence level contours for the non-flat 
$\Lambda$CDM inflation model using various data sets, with the other parameters 
marginalized.\label{fig:tri}}
\end{figure}

\begin{figure}[ht]
\gridline{\fig{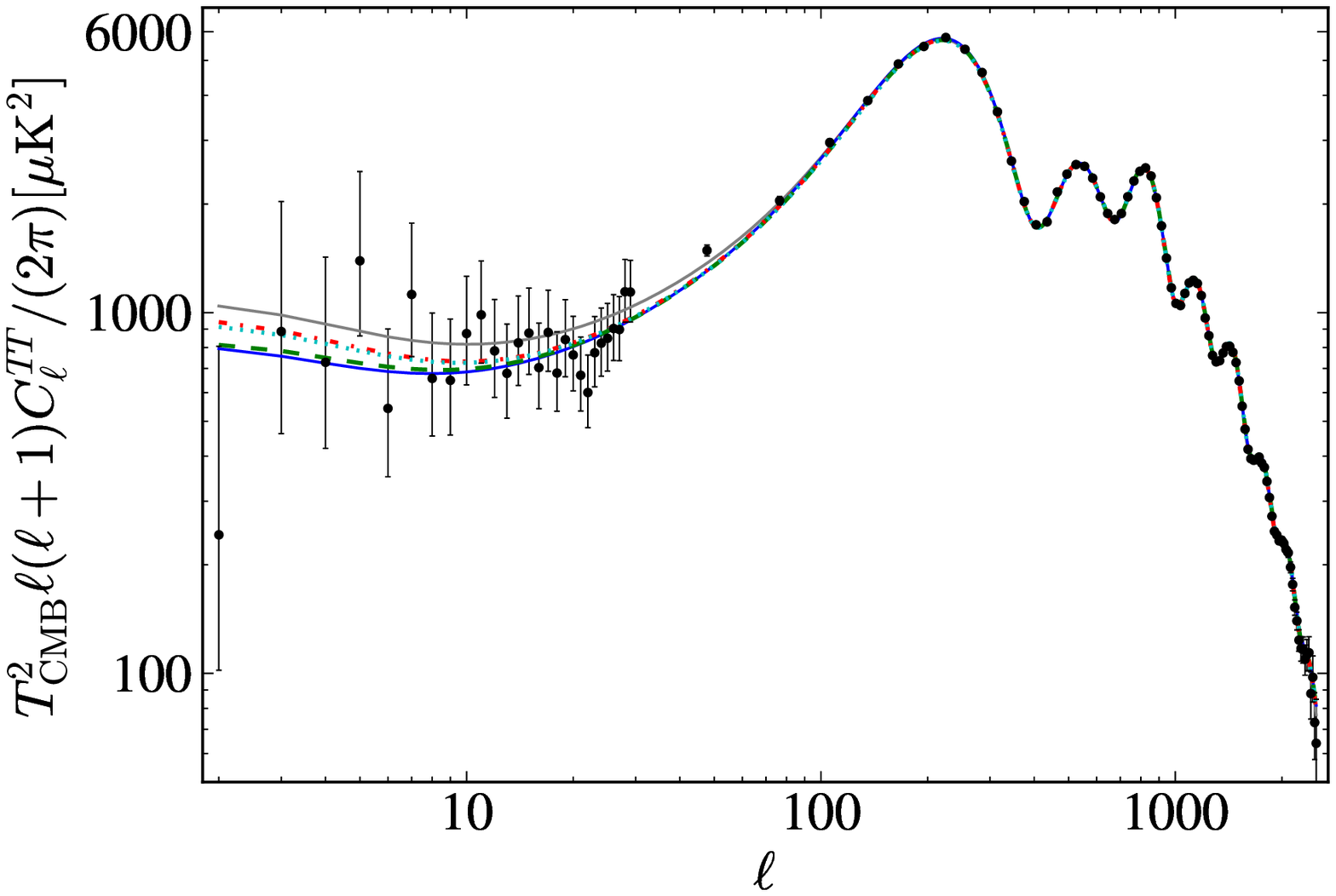}{0.45\textwidth}{(a)}
          \fig{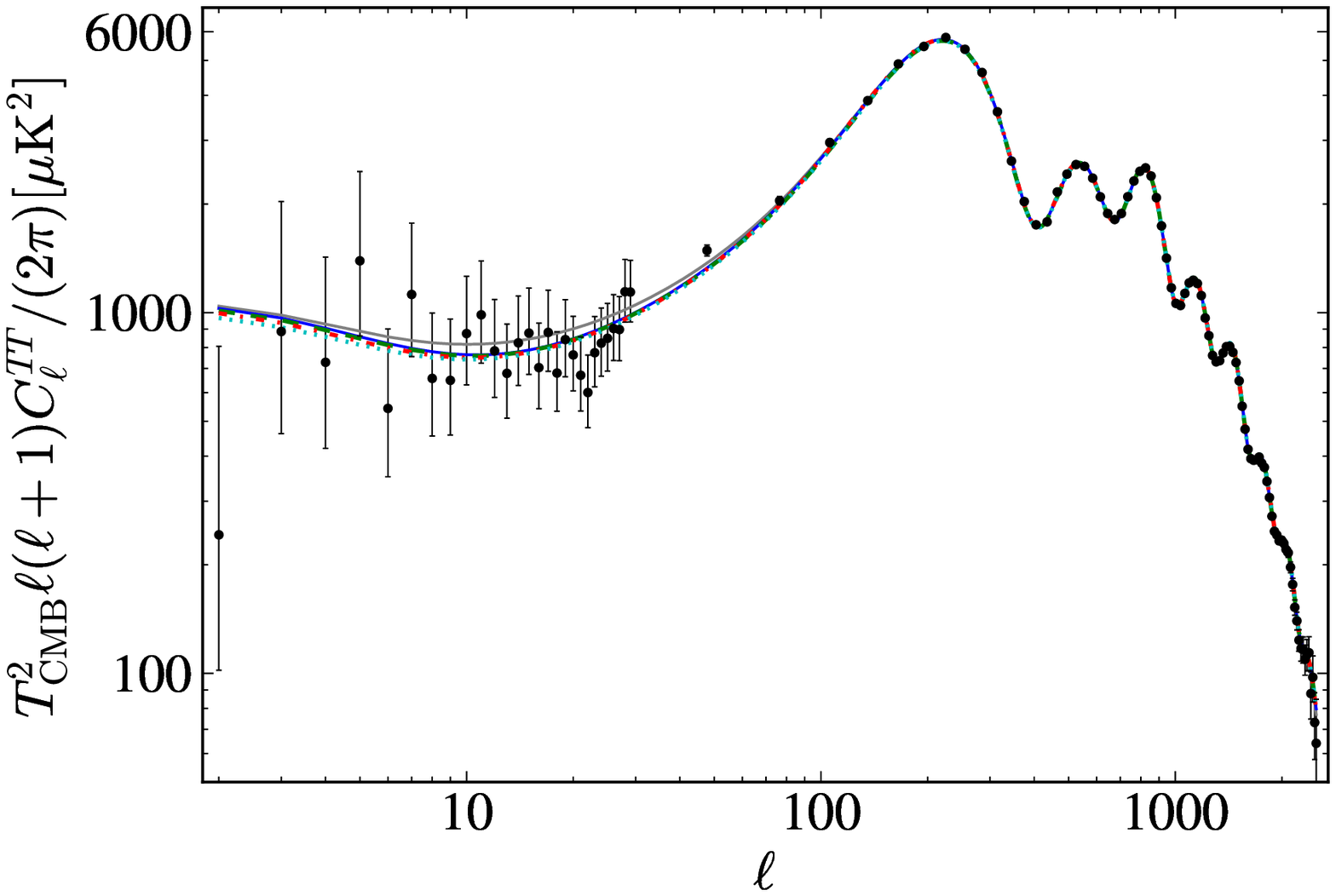}{0.45\textwidth}{(b)}
          }
\gridline{\fig{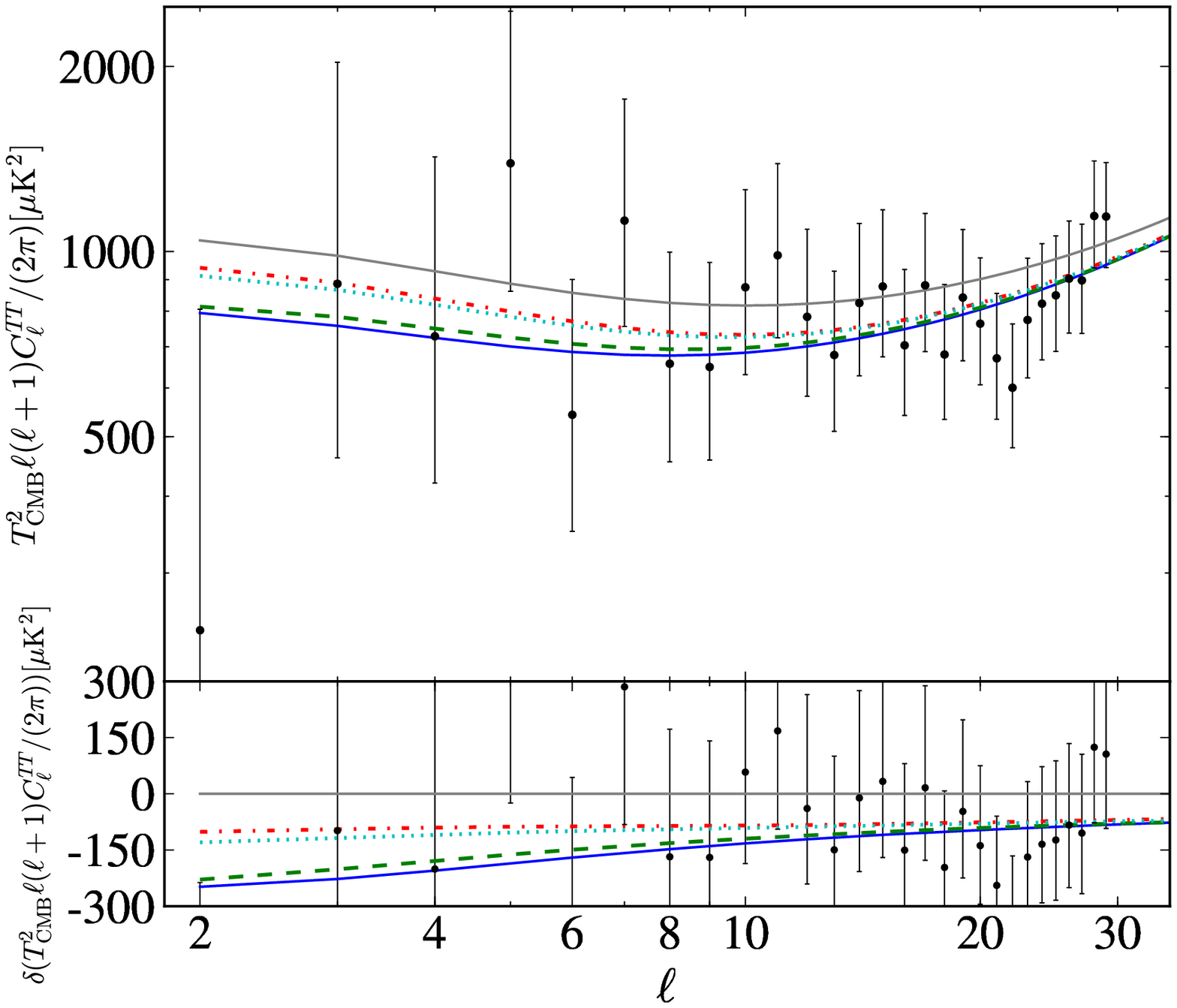}{0.42\textwidth}{(c)}
          \fig{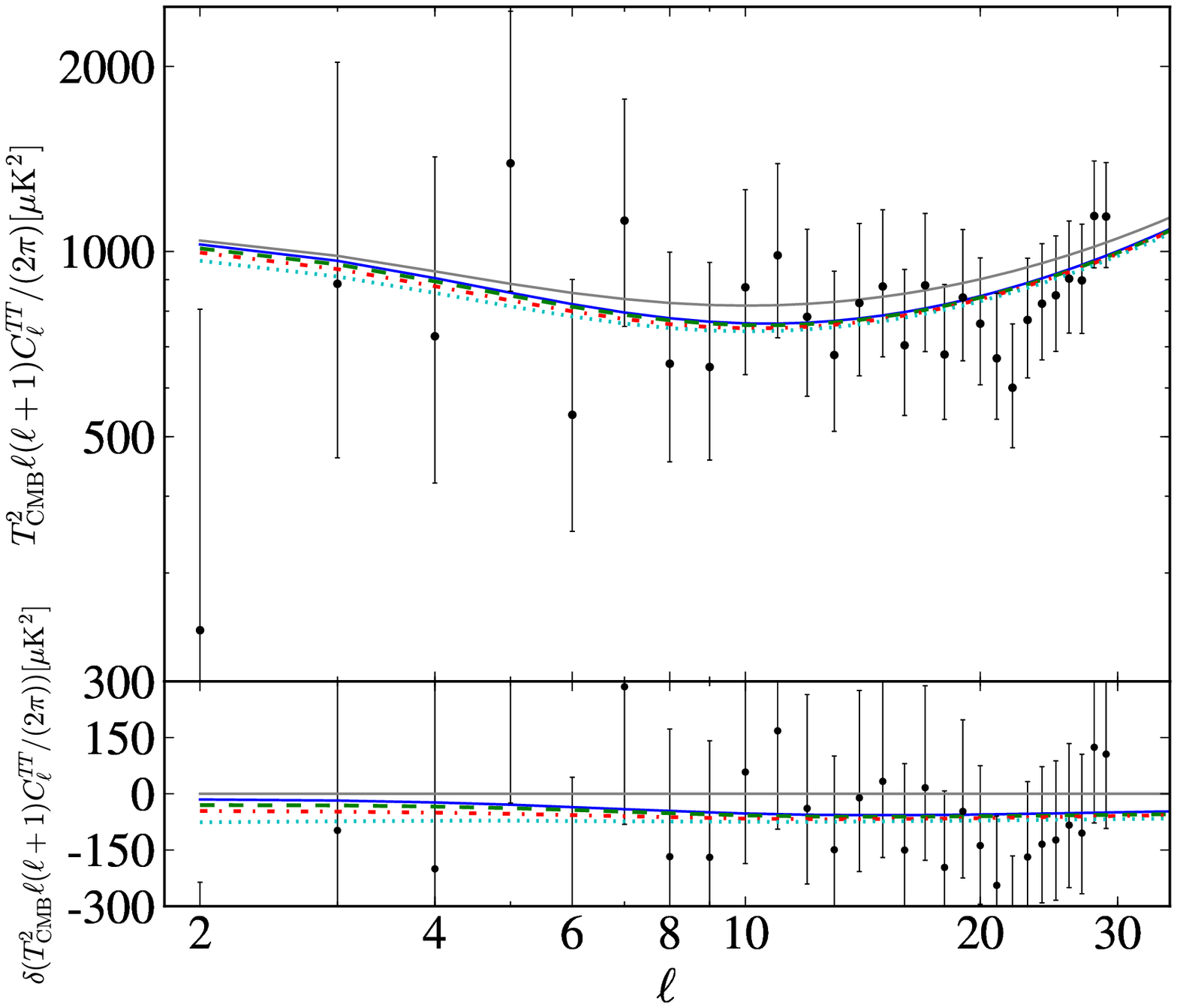}{0.42\textwidth}{(d)}
          }
\gridline{\fig{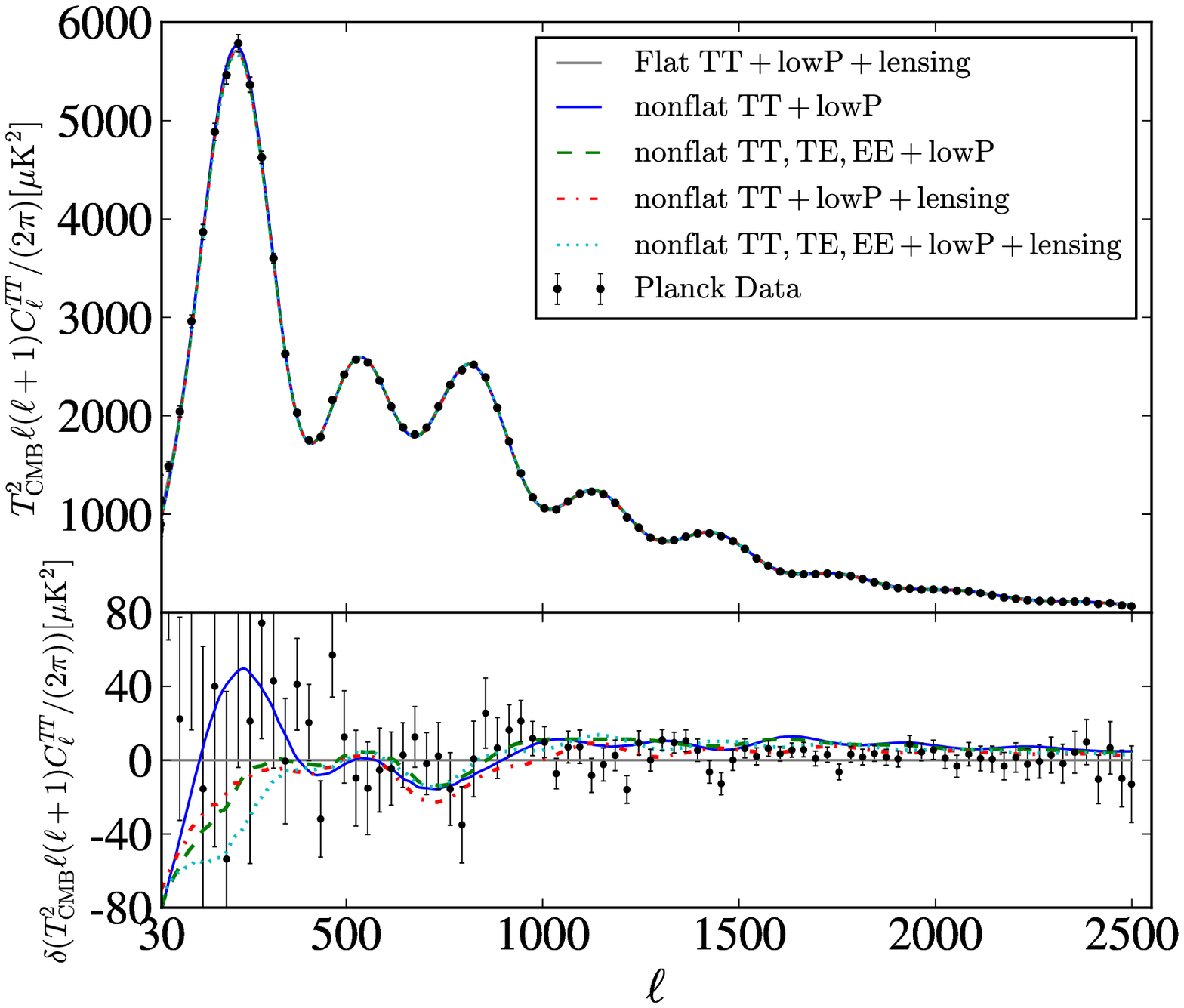}{0.43\textwidth}{(e)}
          \fig{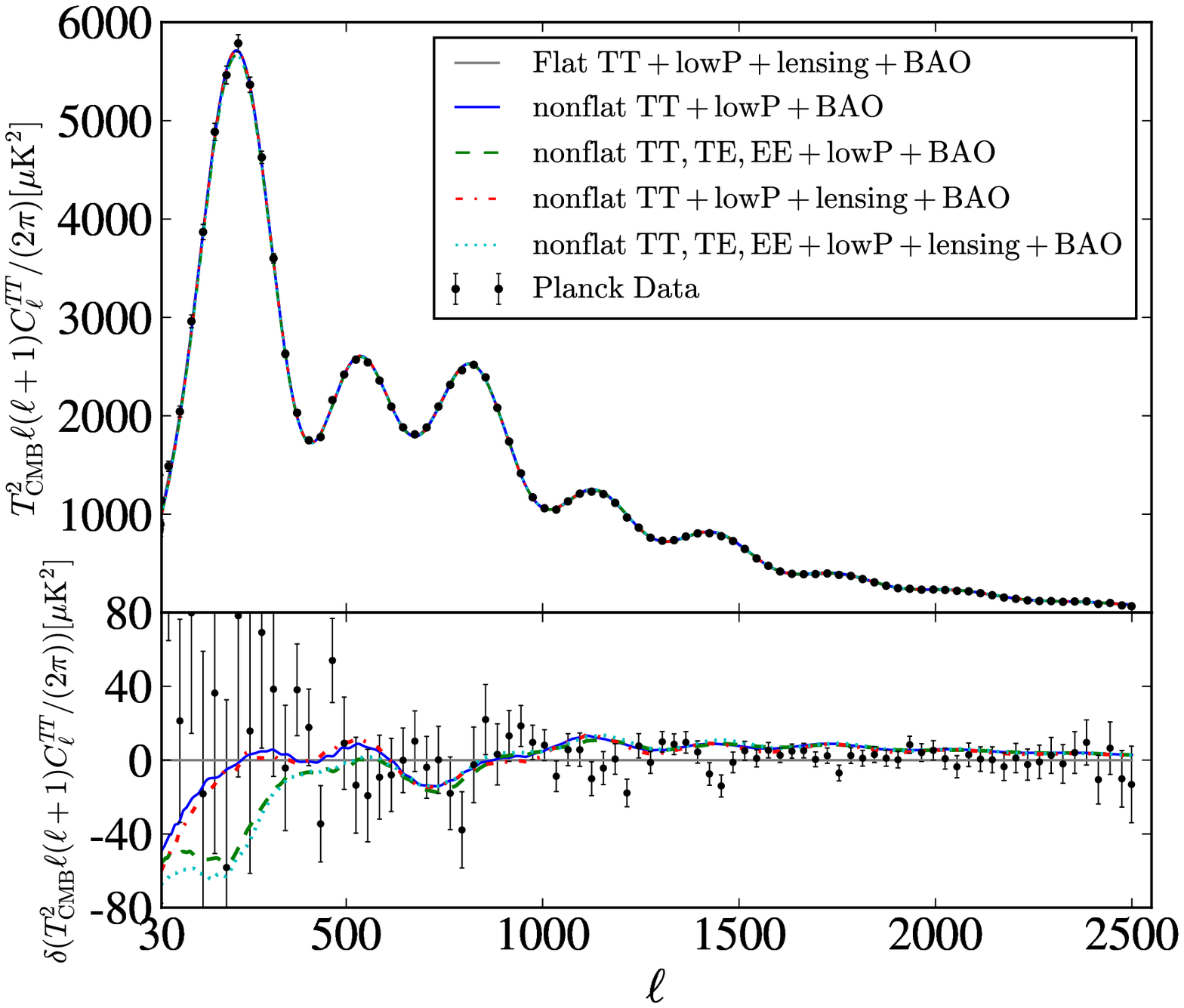}{0.43\textwidth}{(f)}
          }
\caption{The $C_{\ell}$ for the best-fit non-flat $\Lambda$CDM models and the  
spatially-flat tilted $\Lambda$CDM model (gray solid line). Linestyle 
information are in the boxes in the two lowest panels. Planck 2015 data are 
shown as black points with error bars. Left panels (a), (c) and (e) are from 
CMB data alone analyses, while right panels (b), (d) and (f) analyses also
include BAO data. The top panels show the all-$\ell$ region. The middle 
panels show the low-$\ell$ region $C_\ell$ and residuals. The bottom panels show the 
high-$\ell$ region $C_\ell$ and residuals.\label{fig:cls}}
\end{figure}

\begin{figure}[ht]
\plottwo{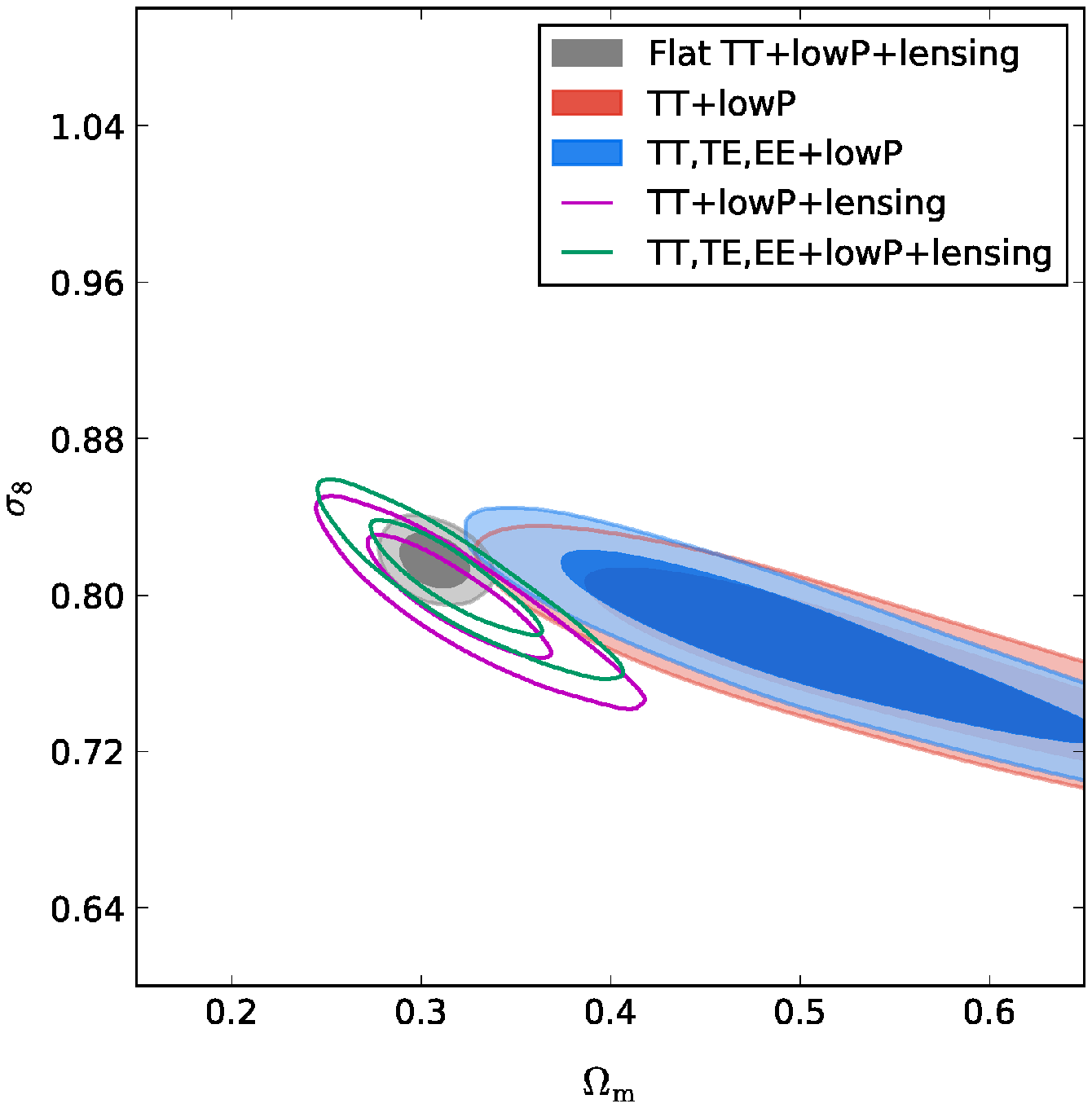}{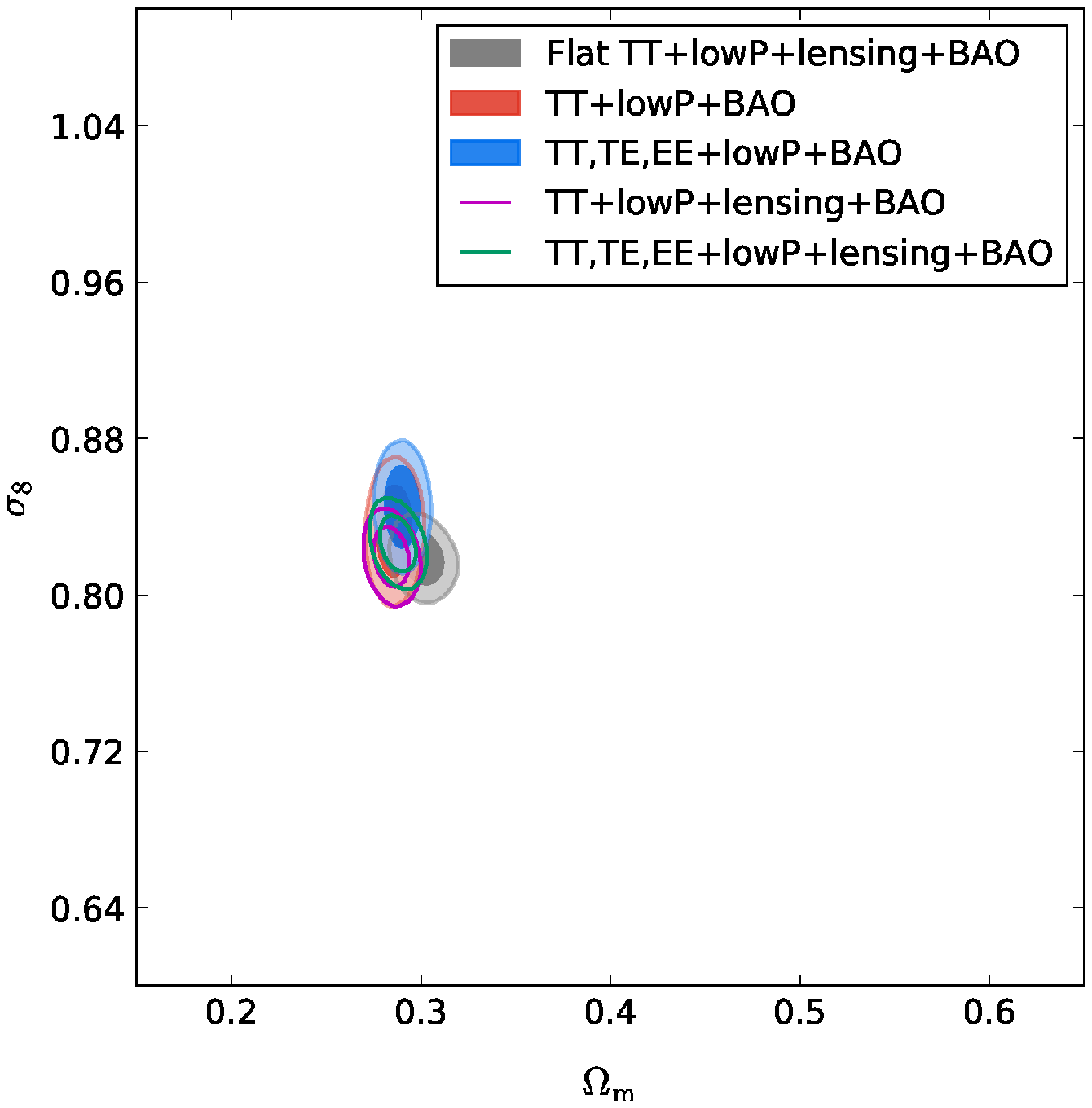}
\caption{$68.27\%$ and $95.45\%$ confidence level contours in 
the $\sigma_8$--$\Omega_{\rm m}$ plane.\label{fig:sigm}}
\end{figure}

In this section, we summarize the results of our parameter estimation 
computations and attempt to judge how well the best-fit 
closed-$\Lambda$CDM inflation model does relative to the best-fit
tilted flat-$\Lambda$CDM inflation model. Table \ref{tab:table1} lists 
central values and 
$68.27\%$ ($1\sigma$) limits on the cosmological parameters from the 
4 different CMB data sets we utilize while Table \ref{tab:table2} lists
the corresponding results from the analyses that also include the BAO 
measurements. Figure \ref{fig:tri} shows two-dimensional constraint contours
and one-dimensional likelihoods determined by marginalizing over all other 
parameters, derived from the 4 CMB anisotropy data sets, both excluding and 
including the BAO data. Figure \ref{fig:cls} shows plots of the CMB 
temperature anisotropy angular power spectra for the best-fit non-flat 
$\Lambda$CDM models determined from the 4 different CMB anisotropy data sets
(as well as one tilted spatially-flat $\Lambda$CDM model), excluding and 
including the BAO data, Figure \ref{fig:sigm} shows $68.27\%$ 
and $95.45\%$ ($2\sigma$) confidence level contours in the 
$\sigma_8$--$\Omega_{\rm m}$ plane, after marginalizing over the other 
parameters, for the non-flat $\Lambda$CDM inflation models as well as for 
one spatially-flat tilted inflation model, for the cases without and with 
the BAO data.

\begin{table*}[ht]
\caption{\label{tab:table1}
68.27\% confidence limits on cosmological parameters of the non-flat $\Lambda$CDM model from CMB data.}
\centering
\begin{tabular}{lcccc}
\hline
\hline
\textrm{Parameter}&
\textrm{TT+lowP}&
\textrm{TT+lowP+lensing}&
\textrm{TT,TE,EE+lowP}&
\textrm{TT,TE,EE+lowP+lensing}\\
\hline
$\Omega_{\rm b}h^2$ & $0.02333\pm 0.00022$ & $0.02304\pm 0.00020$ & $0.02304\pm 0.00015$ & $0.02289\pm 0.00015$\\
$\Omega_{\rm c}h^2$ & $0.1092\pm 0.0011$ & $0.1091\pm 0.0011$ & $0.1108\pm 0.0010$ & $0.1111\pm 0.0009$\\
$100\theta$ & $1.04300\pm 0.00041$ & $1.04306\pm 0.00041$ & $1.04256\pm 0.00030$ & $1.04259\pm 0.00029$\\
$\tau$ & $0.089\pm 0.028$ & $0.101\pm 0.021$ & $0.089\pm 0.026$ & $0.100\pm 0.019$\\
${\rm ln}(10^{10}A_{\rm s})$ & $3.088\pm 0.057$ & $3.108\pm 0.042$ & $3.091\pm 0.053$ & $3.112\pm 0.039$\\
$\Omega_{\rm k}$ & $-0.093\pm 0.037$ & $-0.018\pm 0.008$ & $-0.071\pm 0.028$ & $-0.014\pm 0.008$\\
\hline
$H_0$ [km/s/Mpc] & $48.38\pm 5.77$ & $64.33\pm 3.34$ & $51.14\pm 5.08$ & $65.13\pm 3.14$\\
$\Omega_{\rm m}$ & $0.59\pm 0.13$ & $0.32\pm 0.03$ & $0.53\pm 0.10$ & $0.31\pm 0.03$\\
$\sigma_8$ & $0.751\pm 0.039$ & $0.797\pm 0.022$ & $0.768\pm 0.034$ & $0.808\pm 0.020$\\
\hline
\hline
\end{tabular}
\end{table*}

\begin{table*}[ht]
\caption{\label{tab:table2}
68.27\% confidence limits on cosmological parameters of the non-flat $\Lambda$CDM model from CMB and BAO data.}
\centering
\begin{tabular}{lcccc}
\hline
\hline
\textrm{Parameter}&
\textrm{TT+lowP+BAO}&
\textrm{TT+lowP+lensing+BAO}&
\textrm{TT,TE,EE+lowP+BAO}&
\textrm{TT,TE,EE+lowP+lensing+BAO}\\
\hline
$\Omega_{\rm b}h^2$ & $0.02305\pm 0.00021$ & $0.02302\pm 0.00020$ & $0.02290\pm 0.00015$ & $0.02288\pm 0.00015$\\
$\Omega_{\rm c}h^2$ & $0.1096\pm 0.0011$ & $0.1093\pm 0.0011$ & $0.1114\pm 0.0009$ & $0.1112\pm 0.0009$\\
$100\theta$ & $1.04293\pm 0.00041$ & $1.04302\pm 0.00041$ & $1.04251\pm 0.00029$ & $1.04257\pm 0.00029$\\
$\tau$ & $0.135\pm 0.017$ & $0.120\pm 0.012$ & $0.138\pm 0.016$ & $0.117\pm 0.011$\\
${\rm ln}(10^{10}A_{\rm s})$ & $3.181\pm 0.035$ & $3.150\pm 0.023$ & $3.190\pm 0.033$ & $3.146\pm 0.022$\\
$\Omega_{\rm k}$ & $-0.008\pm 0.002$ & $-0.008\pm 0.002$ & $-0.006\pm 0.002$ & $-0.006\pm 0.002$\\
\hline
$H_0$ [km/s/Mpc] & $68.12\pm 0.75$ & $68.23\pm 0.74$ & $68.05\pm 0.74$ & $68.28\pm 0.74$\\
$\Omega_{\rm m}$ & $0.28\pm 0.01$ & $0.28\pm 0.01$ & $0.29\pm 0.01$ & $0.29\pm 0.01$\\
$\sigma_8$ & $0.819\pm 0.010$ & $0.819\pm 0.010$ & $0.845\pm 0.015$ & $0.826\pm 0.009$\\
\hline
\hline
\end{tabular}
\end{table*}

From the analyses without the BAO data, we find that the spatial curvature 
density parameter is constrained to be
\begin{equation}
\label{eq:omk}
\Omega_{\rm k} = -0.018 {+0.018 \atop -0.020} \ \ (95.45\%,\ \rm TT+lowP+lensing).
\end{equation}
The left hand panels of Fig. \ref{fig:cls} show the CMB temperature anisotropy 
$C_\ell$ of the best-fit non-flat $\Lambda$CDM inflation models for the 4 
different CMB anisotropy data sets. We find that these models fit the 
low-$\ell$ $C_{\ell}$ measurements better than does the spatially-flat tilted 
$\Lambda$CDM case of \citet{Adeetal2016a}, while the higher-$\ell$ $C_{\ell}$ data
are not as well fit by the non-flat models.\footnote{There has been some
discussion of the possibility of slight differences between constraints 
derived from higher-$\ell$ and lower-$\ell$ Planck 2015 CMB anisotropy data 
\citep{Addisonetal2016, Aghanimetal2016} and it might now be useful to 
revisit this issue in the context of the non-flat $\Lambda$CDM inflation model 
we study here.} Figure \ref{fig:Pk} shows that on large scales the fractional 
energy density inhomogeneity power spectrum for the best-fit 
closed-$\Lambda$CDM model is suppressed relative to that of the best-fit 
tilted flat-$\Lambda$CDM model, for the TT + lowP + lensing data. We 
emphasize that while the low-$\ell$ $C_\ell$ of Fig. \ref{fig:cls} depend on 
this small wavenumber part of the power spectrum, other effects, such as 
the usual and integrated Sachs-Wolfe effects, also play an important role 
in determining the $C_\ell$ shape. The left panel of Fig. 
\ref{fig:sigm} shows $\sigma_8$--$\Omega_{\rm m}$ constraint contours,
for the 4 non-flat $\Lambda$CDM models (as well as for one spatially-flat 
tilted $\Lambda$CDM case). When CMB lensing is included, we find that our 
non-flat six parameter $\Lambda$CDM inflation model weakens the tension 
between the CMB observations and the weak lensing data, compare Fig. 3 here 
to Fig. 18 of \citet{Adeetal2016a}.

From the analyses also including the BAO data, we find that the spatial 
curvature density parameter is constrained to be
\begin{equation}
\label{eq:omk2}
\Omega_{\rm k} = -0.008 \pm 0.004\ \ (95.45\%,\ \rm TT+lowP+lensing+BAO).
\end{equation}
Unlike the Planck 2015 results \citep{Adeetal2016a}, our physically-consistent
non-flat $\Lambda$CDM inflation model is not forced to be flat even when we 
include the BAO data in the analysis. Moreover, this case is about 4$\sigma$ 
away from 
flat. The right panels of Fig. \ref{fig:cls} show plots of $C_{\ell}$ for the 
best-fit non-flat $\Lambda$CDM models analyzed using the 4 different CMB 
data sets and including the BAO data. We find that including the BAO data 
does somewhat degrade the fit in the low-$\ell$ region compared with results 
from the analyses without the BAO data. Including the BAO data also worsens the 
$\sigma_8$--$\Omega_{\rm m}$ plane discrepancy between the CMB and weak lensing
constraints, see the right hand panel of Fig. \ref{fig:sigm}.

It is interesting
--- and might even be significant ---
that the $H_0$ and $\Omega_{\rm m}$ constraints listed in the Tables (aside 
from the CMB alone without CMB lensing results of columns 2 and 4 of Table \ref{tab:table1}) are
quite consistent with estimates for these parameters from most other data.
For the density parameter see \citet{ChenRatra2003}.

The most recent median statistics analyses of compilations of $H_0$ 
measurements gives $H_0 = 68 \pm 2.8$ km s${}^{-1}$ Mpc${}^{-1}$ 
\citep{ChenRatra2011}, consistent with earlier values \citep{Gottetal2001, Chenetal2003}. Many more recent $H_0$ determinations from BAO, Type Ia supernovae, 
Hubble parameter, and other measurements are consistent with these results
\citep{Calabreseetal2012, Hinshawetal2013, Sieversetal2013, Aubourgetal2015, LHuillierShafieloo2017, Lukovicetal2016, Chenetal2017}. However, it is well known 
that local measurements of the expansion 
rate give a higher $H_0$. \citet{Freedmanetal2012} report 
$H_0 = 74.3 \pm 2.1$ km s${}^{-1}$ Mpc${}^{-1}$ while \citet{Riessetal2016} 
find $H_0 = 73.24 \pm 1.74$ km s${}^{-1}$ Mpc${}^{-1}$.

\begin{table*}[ht]
\caption{\label{tab:table3}
Minimum $\chi^2$ values for the best-fit closed- (and tilted flat-)$\Lambda$CDM
inflation model.}
\centering
\begin{tabular}{lcc}
\hline
\hline
\textrm{Data sets}& \textrm{$\chi^2$}& \textrm{d.o.f.}\\
\hline
TT+lowP & 11272 (11262) & 189\\
TT+lowP+lensing & 11292 (11272) & 197\\
TT,TE,EE+lowP & 12968 (12936) & 341\\
TT,TE,EE+lowP+lensing & 12991 (12947) & 349\\
TT+lowP+BAO & 11287 (11266) & 191\\
TT+lowP+lensing+BAO & 11298 (11277) & 201\\
TT,TE,EE+lowP+BAO & 12984 (12940) & 345 \\
TT,TE,EE+lowP+lensing+BAO & 12998 (12952) & 353\\
\hline
\hline
\end{tabular}
\end{table*}

\begin{table*}[ht]
\caption{\label{tab:table4}
$\chi^2$ values for the best-fit closed- (and tilted flat-)$\Lambda$CDM
inflation model.}
\centering
\begin{tabular}{lcccc}
\hline
\hline
\textrm{Data}&
\textrm{TT+lowP}&
\textrm{TT+lowP+lensing}&
\textrm{TT,TE,EE+lowP}&
\textrm{TT,TE,EE+lowP+lensing}\\
\hline
CMB low-$\ell$ & 121.49 (129.83) & 125.39 (126.06) & 122.79 (130.75) & 121.78 (126.52)\\
CMB high-$\ell$ & 147.90 (77.13) & 123.24 (82.46) & 310.88 (212.67) & 314.11 (224.36)\\
CMB all-$\ell$ & 269.39 (206.97) & 258.46 (218.43) & 433.67 (343.42) & 445.91 (360.86)\\
CMB lensing & --- & 9.82 (9.90) & --- & 10.02 (9.98)\\
\hline
d.o.f.\ & 189 & 197 & 341 & 349 \\
\hline
\hline
\end{tabular}
\end{table*}

\begin{table*}[ht]
\caption{\label{tab:table5}
$\chi^2$ values for the best-fit closed- (and tilted flat-)$\Lambda$CDM
inflation model.}
\centering
\begin{tabular}{lcccc}
\hline
\hline
\textrm{Data}&
\textrm{TT+lowP+BAO}&
\textrm{TT+lowP+lensing+BAO}&
\textrm{TT,TE,EE+lowP+BAO}&
\textrm{TT,TE,EE+lowP+lensing+BAO}\\
\hline
CMB low-$\ell$ & 142.89 (130.06) & 130.11 (126.00) & 137.88 (132.56) & 124.44 (126.56)\\
CMB high-$\ell$ & 135.05 (73.64) & 133.63 (80.82) & 307.86 (211.10) & 324.95 (222.60)\\
CMB all-$\ell$ & 284.31 (207.73) & 278.20 (220.31) & 451.32 (347.93) & 464.19 (363.11)\\
CMB lensing & --- & 9.86 (9.92) & --- & 10.01 (9.96)\\
BAO & 6.35 (4.03) & 4.59 (3.58) & 5.57 (4.27) & 4.78 (3.99)\\
\hline
d.o.f.\ & 191 & 201 & 345 & 353\\ 
\hline
\hline
\end{tabular}
\end{table*}

Additionally, we emphasize that many analyses based on a variety of different observations also do not rule out non-flat dark energy models
\citep{Farooqetal2015, Saponeetal2014, Lietal2014, Caietal2016, Chenetal2016, YuWang2016, LHuillierShafieloo2017, Farooqetal2017, Lietal2016, WeiWu2017, Ranaetal2017, Yuetal2018, Mitraetal2018, Ryanetal2018}.

It is also important to understand how well the best-fit closed-$\Lambda$CDM 
inflation model does relative to the best-fit tilted flat-$\Lambda$CDM model in fitting the data.
Unfortunately we are unable to resolve this in a quantitative manner,
although qualitatively, overall, the best-fit closed-$\Lambda$CDM model 
does not do as well as the best-fit tilted flat-$\Lambda$CDM 
model.\footnote{We note that the tilted flat-$\Lambda$CDM model and
the closed-$\Lambda$CDM model are not nested. Rather, the best-fit versions
of each of these six parameter models form two distinct local likelihood 
maxima in a larger seven parameter model space.} However, it
appears that it does reasonably well enough to warrant a more thorough, 
quantitative, study of this issue. 

Table \ref{tab:table3} lists the minimum $\chi^2 = -2 {\rm ln} 
({\rm L_{\rm max}})$ determined from the maximum value of the 
likelihood,\footnote{This is the $\chi_{\rm eff}^2$ of 
Planck 2015 Results: \\Cosmological Parameter Tables at
wiki.cosmos.esa.int/planckpla2015/images/f/f7/Baseline{\_}params{\_}table{\_}2015{\_}limit68.pdf.}
for 
the eight data sets we study, for both the closed-$\Lambda$CDM and tilted
flat-$\Lambda$CDM inflation models, as well as the number of (binned data) 
degrees of freedom (d.o.f.). The d.o.f.\ are determined from combinations of
112 low-$\ell$ TT + lowP, 83 high-$\ell$ TT, 132 high-$\ell$ TE and EE, 8 
lensing CMB (binned) measurements, 4 BAO measurements, and 6 model parameters. 
Almost certainly the large $\chi^2$ values are the result of the many nuisance 
parameters that have been marginalized over,\footnote{We have been unable to 
determine definite quantitative information about this.} 
as the tilted flat-$\Lambda$CDM model is said to be a good fit to the data.
From this table we see that $\Delta \chi^2 = 20 (21)$ for the 
closed-$\Lambda$CDM inflation model, relative to the tilted flat-$\Lambda$CDM, 
for the 197 (201) d.o.f.\ for the TT + low P + lensing (+ BAO) data 
combination. While this might make the closed-$\Lambda$CDM model much 
less probable, one can see from the residual panels of Fig. \ref{fig:cls} (e)
\& (f) that this $\Delta \chi^2$ is apparently caused by
many small deviations,
and not by a few significant outliers. This then allows for the possibility 
that a slight increase in the error bars or a mild nonGaussinity in the errors
could raise the model  probabilities.    

While there are correlations in the data,
it is also instructive to consider 
a standard goodness of fit $\chi^2$ that only makes use of the diagonal 
elements of the correlation matrix. These are listed in Tables \ref{tab:table4}
\& \ref{tab:table5} for the eight data sets we study and for both the 
closed-$\Lambda$CDM and tilted flat-$\Lambda$CDM inflation models. From Table
\ref{tab:table4} for the TT + low P + lensing data, we see that the $\chi^2$ 
per d.o.f.\ is 268/197 (228/197) for the closed-$\Lambda$CDM (tilted 
flat-$\Lambda$CDM) inflation model, while from Table \ref{tab:table5}, when 
BAO data is added to the mix, these become 293/201 (234/201). Again, 
while the closed-$\Lambda$CDM model is less favored than the tilted 
flat-$\Lambda$CDM case,
it is not straightforward to assess the quantitative significance of this.
In addition to the points mentioned at the end of the 
previous paragraph, in this analysis we also ignore all the off-diagonal 
information in the correlation matrix, so it is meaningless to compute 
standard probabilities from such $\chi^2$'s. All in all, we believe that our 
results call for a  more thorough analysis of the closed-$\Lambda$CDM 
inflation model.

\section{Conclusion}

We present Planck 2015 CMB data constraints on the physically consistent 
six parameter non-flat $\Lambda$CDM model with inflation-generated 
non-power-law energy density inhomogeneity power spectrum. Unlike the case for 
the seven parameter non-flat tilted $\Lambda$CDM model with power-law power 
spectrum used in \citet{Adeetal2016a}, we discover that CMB anisotropy data 
do not force spatial curvature to vanish in our non-flat inflation model. 
Spatial 
curvature contributes about 2 \% to the present energy budget of the closed 
model that best fits the Planck TT + lowP + lensing data. This model is more 
consistent with the low-$\ell$ $C_{\ell}$ observations and the weak lensing 
$\sigma_8$ constraints than is the best fit spatially-flat tilted $\Lambda$CDM,
but it does worse at fitting the higher-$\ell$ ${C_{\ell}}$ measurements.

It might be useful to revisit the issue of possible small differences in 
the constraints derived from higher-$\ell$ and lower-$\ell$ Planck 2015 CMB 
anisotropy data, by using the non-flat $\Lambda$CDM model we have used here.
Also useful would be a method for quantitatively assessing how well the 
best-fit tilted spatially-flat $\Lambda$CDM model and the best-fit non-flat 
$\Lambda$CDM model fit the CMB anisotropy (and other) data.

Unlike the analysis for the seven parameter non-flat tilted $\Lambda$CDM model 
in \citet{Adeetal2016a}, adding the BAO data still does not force our 
physically consistent six parameter non-flat $\Lambda$CDM model to be flat,
in fact $\Omega_{\rm k} =  -0.008 \pm 0.004$ at 2$\sigma$ and is about 
4$\sigma$ away from flat. In this case the improved agreement with the 
low-$\ell$ $C_{\ell}$ observations and the weak lensing $\sigma_8$ are not 
as good compared with the results from the analyses using only the Planck 
2015 CMB data. However, the BAO and CMB data are from very disparate redshifts 
and it is possible that a better model for the intervening epoch or an improved 
understanding of one or both sets of measurements might alter this result.
Our main motivation in utilizing the BAO data here was to check whether,
when these data are combined with the CMB anisotropy data, they force the 
model to be flat. We emphasize that our non-flat $\Lambda$CDM inflation model 
is not forced to be flat even when the BAO data are added to the 
mix.\footnote{ Our results here have been confirmed \citep{ParkRatra2018a, ParkRatra2018b} using an independent code (based on CAMB/COSMOMC; we use CLASS and Monte Python here) for analyses of the Planck 2015 CMB data in conjunction with significantly more non-CMB data than we use here, resulting in a significant increase in the evidence for non-flatness in the non-flat $\Lambda$CDM inflation model, to 5.2$\sigma$.}   

In both cases (with and without BAO measurements) CMB anisotropy data 
constraints on $H_0$ and $\Omega_{\rm m}$ are consistent with most other 
constraints on these two parameters.

We have recently completed similar analyses, with similar conclusions,
of the seven parameter non-flat XCDM inflation model \citep{Oobaetal2017a, ParkRatra2018b}
and the seven parameter non-flat $\phi$CDM inflation model 
\citep{Oobaetal2017b, ParkRatra2018c}.

Perhaps a small spatial curvature contribution, of order a few percent, can 
improve the currently popular spatially-flat standard $\Lambda$CDM model.
However, a more thorough analysis of the non-flat $\Lambda$CDM inflation 
model is needed to establish if it is viable and if it can help resolve some of 
the low-$\ell$ $C_{\ell}$ issues as well as possibly the $\sigma_8$ power issues.

\section*{Acknowledgments}

We acknowledge valuable discussions with M. Bucher, K. Ganga, L. Page, and 
J. Peebles. This work is supported by Grants-in-Aid for Scientific 
Research from JSPS (Nos.\ 16J05446 (J.O.) and 15H05890 (N.S.)). B.R.\ is 
supported in part by DOE grant DE-SC0011840.


\end{document}